\documentclass[journal=jpclcd,manuscript=letter,layout=onecolumn]{achemso}
\setkeys{acs}{usetitle = true}
\usepackage{booktabs}
\usepackage{txfonts}

\title{Coherent exciton dynamics in the presence of underdamped vibrations}
\author{Arend~G.~Dijkstra}
\affiliation{Department of Chemistry, Massachusetts Institute of Technology, 77 Massachusetts Avenue, Cambridge, MA 02139, United States of America}
\author{Chen Wang}
\affiliation{Singapore-MIT Alliance for Research and Technology, 1 CREATE Way, Singapore 138602, Singapore}
\author{Jianshu Cao}
\affiliation{Department of Chemistry, Massachusetts Institute of Technology, 77 Massachusetts Avenue, Cambridge, MA 02139, United States of America}
\alsoaffiliation{Singapore-MIT Alliance for Research and Technology, 1 CREATE Way, Singapore 138602, Singapore}
\email{jianshu@mit.edu}
\author{Graham~R.~Fleming}
\affiliation{Physical Biosciences Division, Lawrence Berkeley National Laboratory, Berkeley, CA 94720, United States of America}
\alsoaffiliation{Department of Chemistry, University of California, Berkeley, CA 94720, United States of America}


\begin{document}

\begin{abstract}
Recent ultrafast optical experiments show that excitons in large biological light-harvesting complexes are coupled to molecular vibration modes.
These high-frequency vibrations will not only affect the optical response, but also drive the exciton transport. Here,  using a model dimer system, the frequency of the underdamped vibration is shown to have a strong effect on the exciton dynamics such that quantum coherent oscillations in the system can be present even in the case of strong noise. Two mechanisms are identified to be responsible for the enhanced transport efficiency: critical damping due to the tunable effective strength of the coupling to the bath, and resonance coupling where the vibrational frequency coincides with the energy gap in the system.  The interplay of these two mechanisms determines parameters responsible for the most efficient transport, and these optimal control parameters are comparable to those in realistic light-harvesting complexes.  Interestingly, oscillations in the excitonic coherence at resonance are suppressed in comparison to the case of an off-resonant vibration.
\end{abstract}
\maketitle

\begin{figure}[t]
\includegraphics[width=4.5cm]{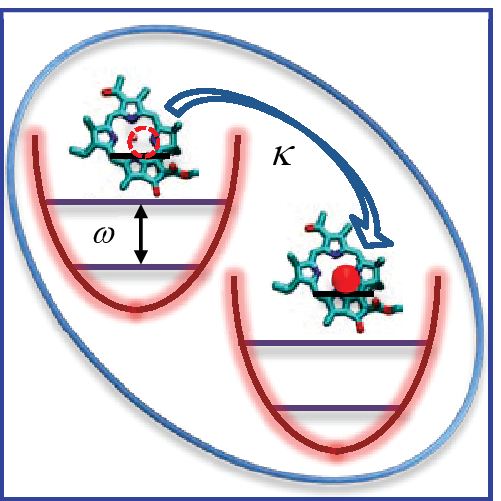}
\end{figure}

\emph{Introduction}--
The effect of intramolecular vibrations on energy transport in large biological assemblies became increasingly intriguing,
for recent optical experiments have been interpreted as a sign of the coupling of vibrations to electronic excitations (excitons) in light-harvesting antennae\cite{Tiwari.2013.pnas.110.1203}, 
and the reaction center of photosynthetic complexes\cite{Fuller.2013.arxiv,Romero.2014.naturephys.10.676}.
These systems consist of closely spaced chromophore molecules.
Coherent interactions of the
transition dipoles of these molecules lead to elementary excitations
that extend over a number of molecules in the form of Frenkel excitons. The existence of these states enables quantum mechanical, wave-like transport through the systems\cite{Engel.2007.nature.446.782, Collini.2009.science.323.369,Collini.2010.nature.463.644}.
The importance of the interaction of such excitons with their environments is well known, and the interplay of coherent coupling between molecules and noise originating from the environment has been shown to lead to optimal transport\cite{Plenio.2008.njp.10.113019,Wu.2010.njp.12.105012}.

To model the influence of the environment, it is often assumed that its effect is fast compared to the typical time scales of the system such that the environment can be modeled as white noise.
However, recent optical experiments detect rapid dynamics of electronic excitations on the time scale of
tens to hundreds femtoseconds in molecular aggregates\cite{Dijkstra.2008.jcp.128.164511}, light-harvesting systems of bacteria\cite{Engel.2007.nature.446.782, Collini.2010.nature.463.644},
plants\cite{Schlau.2012.naturechem.4.389} and conjugated polymers\cite{Collini.2009.science.323.369}.
Then, the environment cannot be considered fast on this time scale. In particular, in many small organic and biological molecules,
the coupling of excitations to vibrations is essential\cite{Adolphs.2006.biophysj.91.2778,Olbrich.2011.jpcb.115.8609, Valleau.2012.jcp.137.224103, Gelin.2011.pre.84.041139, Christensson.2012.jpcb.116.7449,Nalbach.2011.pre.84.041926, Chin.2013.nphys.9.113, Tiwari.2013.pnas.110.1203}.
Not much is known about the transport properties of a system coupled to these underdamped vibrations.
In this paper, we focus on the exciton dynamics induced by underdamped vibrations,
and our method is applicable to baths as Gaussian colored noise.
Initial analysis beyond the white noise limit has included overdamped vibrations.
Recently, it has been realized that underdamped vibrations or vibronic states with a mixed exciton character can also lead to long-lived oscillations during the waiting time\cite{Sharp.2013.jcp.139.144304}. Much work has been devoted to the explanation of these oscillations and to the description of exciton coherence\cite{Kreisbeck.2012.jpcl.3.2828, Chin.2013.nphys.9.113, Tiwari.2013.pnas.110.1203}. 
Exciton dynamics in the presence of underdamped vibrations has recently attracted much attention as a way to 
explain the quantum beats observed in two-dimensional optical experiments\cite{Engel.2007.nature.446.782}, 
and resonant vibrations have been proposed to drive exciton coherences in the system\cite{Christensson.2012.jpcb.116.7449, Tiwari.2013.pnas.110.1203, Chin.2013.nphys.9.113}. It is not yet clear, however, which parameters can optimize transport in a system coupled to underdamped vibrations. Resonance between an energy gap of the system and a vibrational mode is a possible mechanism but not the only one\cite{Rey.2013.jpcl.4.903}; an equally important mechanism is the critical damping, where the exciton dynamics undergo a transition from underdamped to overdamped oscillations.

In this work, we address the effect of an underdamped vibration on energy transport. If vibrations couple strongly to excitons, as suggested by the recent explanations of long-lived coherences, they are also expected to influence the population dynamics underlying energy transport. In order to clearly bring out the essential physics of the interplay between excitonic coherence and effect of an underdamped vibration, we study a prototype model: an electronic dimer coupled to a single underdamped vibration. The transport efficiency in such a system depends strongly on the coupling to the vibration and is found to be highly non-trivial.

\emph{Model system}--
The prototype model that describes exciton delocalization is an electronic dimer, with two molecules labeled $1$ and $2$,
and extension of this model to larger systems is straightforward.
Each molecule is modeled as a two-level system with a common ground state and an excited state. The excitation energy of molecule  $1$ ($2$) is denoted $\epsilon_1$ ($\epsilon_2$) and the coherent interaction between molecules as $-J$. The full Hamiltonian is given in terms of the creation and annihilation operators $\hat{c}^{\dag}_{1(2)}$ and $\hat{c}_{1(2)}$ by
\begin{eqnarray}~\label{ham1}
  \hat{H}(t)&=&[\epsilon_1 + \delta \epsilon_1({X}_1(t))] \hat{c}^{\dag}_{1}\hat{c}_{1} + [\epsilon_2 + \delta \epsilon_2({X}_2(t))] \hat{c}^{\dag}_{2}\hat{c}_2  \nonumber \\
       &-& J ( \hat{c}^{\dag}_{1}\hat{c}_{2} + \hat{c}^{\dag}_{2}\hat{c}_{1}),
\end{eqnarray}
where the system is on-site coupled to the baths.
The coordinates of the environment $X_n$ comprise all degrees of freedom not included in the tight-binding Hamiltonian, and, in particular, vibrations. In principle, these $X_n$'s are operators that must be described by the rules of quantum mechanics. Their correlation function $\langle X_n(t) X_n(0)\rangle$ is a complex quantity, with its real and imaginary parts balanced by the fluctuation-dissipation theorem. However, in the spirit of stochastic modeling, we first assume that  $X_n(t)$ is a real function of time. The energy fluctuations $\delta \epsilon$ are then random variables which follow a specific correlation function. This approximation is valid when the temperature is large compared with the bandwidth of the system.
In our numerical simulations, we will consider a dimer with the same site energies $\epsilon_1 = \epsilon_2$ (homodimer) as well as a dimer with different
site energies (heterodimer), where the offset in site energies is $\Delta = \epsilon_1 - \epsilon_2$.
Furthermore, we will use the excitonic coupling $J$ as the energy unit.
At the end of our paper, we will show that the essential physics obtained from our model with classical fluctuations is retained,
when the environment is modeled 
quantum mechanically.

Although this model is standard, most studies assume that the fluctuating excitation energies are either
stationary random variables\cite{Malyshev.2007.prl.98.087401}, or as white noise which can be modeled in the Markov approximation\cite{Plenio.2008.njp.10.113019, Wu.2013.prl.110.200402}. Here, we consider general Gaussian colored noise with the correlation function
\begin{equation} \label{cfdef}
L(t) = \langle \delta \epsilon_n(t) \delta \epsilon_n(0)\rangle.
\end{equation}
We assume that the fluctuations in the site energies are uncorrelated and the correlation functions are identical for every site, although these assumptions can easily be relaxed.
The correlation function for damped vibrations is given by
\begin{equation} \label{cf}
L(t) = \frac{\Gamma_0}{\tau} \cos (\omega t) e^{-|t| / \tau}.
\end{equation}
The two parameters $\Gamma_0/\tau$ and $\omega$ describe the amplitude of fluctuations and the frequency of vibration. The parameter $\tau$ models the memory time scale of the bath, which is equal to the inverse damping rate of the vibration. For $\omega=0$ the correlation function in Eq.~(\ref{cf}) describes overdamped vibrations and can be derived from a Langevin equation driven by a white noise\cite{Anderson.1954.jpsj.9.316, Kubo.1954.jpsj.9.935}. In this case, $1/\tau$ is known as the Debye frequency. The presence of memory (i.e., $\tau > 0$) can increase the coherence in the system \cite{Ishizaki.2009.pnas.106.17255} and hence the exciton diffusion efficiency. Here, we focus on the effect of nonzero $\omega$ to describe underdamped vibrations.

We are interested in the exciton dynamics with one excitation present.
It is known that from Eq.~(\ref{ham1}) the system Hamiltonian and system-bath interaction do not commute.
Hence, there is no analytical solution for the dynamics beyond the white noise limit.
The two-state dynamics is therefore found numerically by solving the time-dependent Schr\"odinger equation.
The calculation is repeated for many realizations of the random noise and the density matrix is averaged over all the realizations. All simulations in this paper were averaged over $10^4$ noise trajectories.
To do this, we first generate trajectories of random processes $\{\delta\epsilon_n(t_i)\}$ for the $n$th site, with discrete time series $\{t_i\}$.
It is practically obtained by applying a linear transformation $\delta\epsilon_n(t_i)=\sum_{k}\mathcal{A}_{ik}\delta\eta_n(t_k)$,
where $\{\delta\eta_n(t_k)\}$ are sequences of white noise, with correlation ${\langle}\delta\eta_n(t_k)\delta\eta_n(t_l){\rangle}=\delta_{k,l}$.
Moreover, the transformation matrix $\mathcal{A}$ is constrained by the correlation function $L(t)$,
and is computed by Cholesky decomposition of the covariance matrix\cite{Press.2002.book}, specified as $L(t_{ij})=(\mathcal{A}\mathcal{A}^{T})_{ij}$.
Hence, the noise correlation function is recovered as
${\langle}\delta\epsilon_{n}(t_i)\delta\epsilon_{n}(t_j){\rangle}
=\sum_{l,k}\mathcal{A}_{ik}\mathcal{A}_{jl}{\langle}\delta\eta_{n}(t_k)\delta\eta_{n}(t_l){\rangle}
=\sum_k\mathcal{A}_{ik}(\mathcal{A}^{T})_{kj}=L(t_{ij})$.
Populations and coherences at each time step
are therefore found by storing the components of the density matrix $\rho_{nm}(t) = {\langle}\psi_n(t) \psi_m^*(t){\rangle}$.

\emph{Critical damping}--Let us first briefly consider the white noise case. The correlation function is then given by $L(t) = \Gamma \delta(t)$, where the dephasing rate $\Gamma$ quantifies the strength of the noise. In this case, the average over the noise can be performed analytically and leads to the Haken-Strobl-Reineker model of exciton transport. The equation of motion for the reduced density matrix is found to be
$\frac{d}{dt}{\hat{\rho}}(t) = -i [\hat{H}_S, \hat{\rho}(t)] + \hat{L}_d [\hat{\rho}(t)]$
with $\langle n|\hat{L}_d[\hat{\rho}(t)]|m\rangle = - \Gamma (1-\delta_{nm}) \langle n|\hat{\rho}(t) |m\rangle$ in the site basis\cite{Moix.2013}, where $\hat{H}_S$ is the system Hamiltonian without coupling to the bath, and $|n{\rangle}$ is the site state. The behavior of the system can be tuned by varying $\Gamma$.
For small $\Gamma$ ($\Gamma < \Gamma_\mathrm{critical}$) the population dynamics exhibit oscillations, such that irreversible transfer of population from one site to the other is slow. We will refer to this situation as the underdamped regime. For large $\Gamma$, the transfer is incoherent (overdamped regime) and slows down with increasing $\Gamma$.
In between, the critical value of $\Gamma$ optimizes the transfer (critical damping), which has been postulated as optimal energy transfer in light-harvesting systems. Although an optimal value of $\Gamma$ exists, there is a broad plateau of values close to the maximum that lead to near-optimum transport\cite{Wu.2010.njp.12.105012}.
The coherent oscillations disappear at the critical value of $\Gamma = 4 J$ based on the eigenvalues of the Liouville operator,
which is analytically found in the homodimer. These eigenvalues are $0$, $-\Gamma$ and $\frac{1}{2}(-\Gamma \pm \sqrt{\Gamma^2-16 J^2})$. Critical damping corresponds to the crossover from purely real to complex eigenvalues, which occurs when the latter two eigenvalues become equal at $\Gamma = 4 J$.
In the case of the heterodimer, though not shown here, it is numerically found that for $\Delta > 2J$, the critical value is best understood by the energy gap of the system,
$\Gamma_\mathrm{critical} {\approx} \sqrt{\Delta^2 + 4 J^2} {\approx} \Delta$.

For colored noise with a more general correlation function, a similar phenomenon is expected. In the Markovian limit, when the bath time scale is much faster than the system dynamics, the effect of the bath can be understood with an effective damping rate $\Gamma_\mathrm{eff} = \int_0^\infty \mathrm{d}t L(t)$. Thus, a crossover from the underdamped to the overdamped regime is expected when the integral of the correlation function increases. This can be achieved by increasing the strength of the fluctuations, given by $L(0)$, by changing the effective memory time of the bath, or by changing the shape of the correlation function.
Although the argument of effective model for the damping does not always hold outside the Markovian limit, it provides qualitative insight.

\begin{figure*}[t]
 \includegraphics{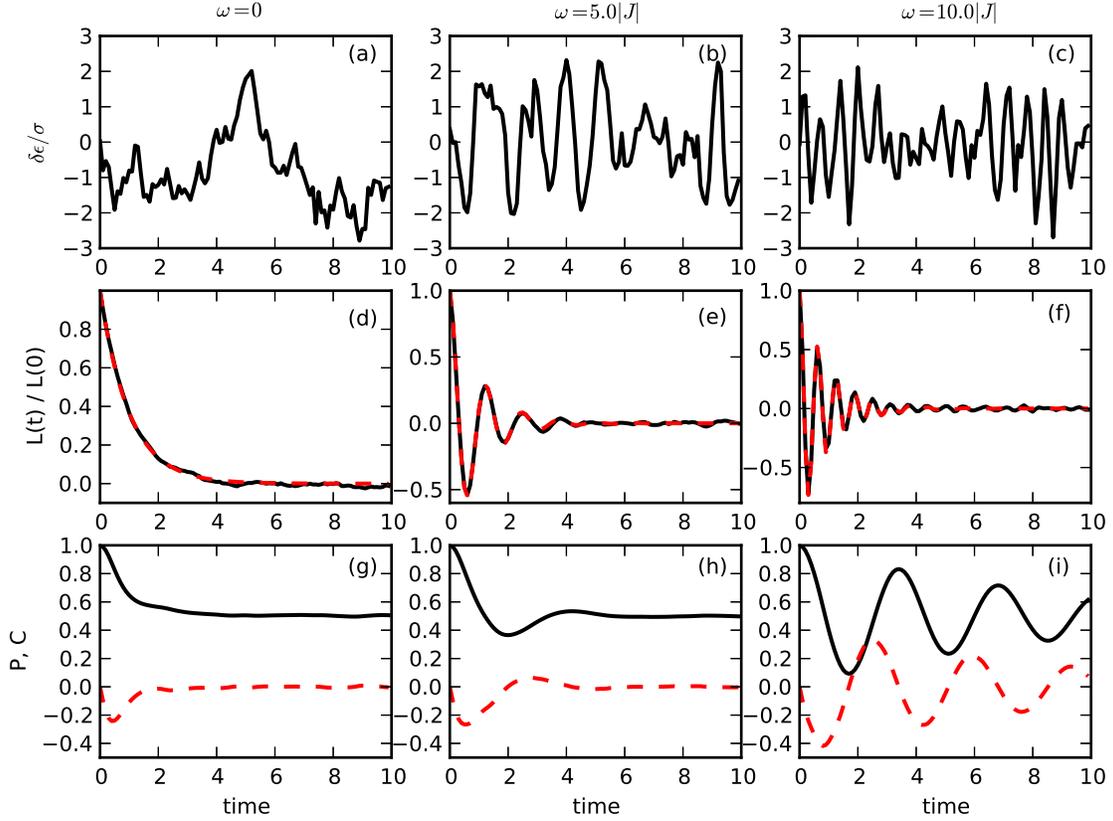}
\caption{\label{fig:dimerseqcfpop} Panels (a) - (c) show realizations of the noise for $\omega = 0.0, 5.0, 10.0 J$, respectively. Panels (d) - (f) show the chosen correlation function as well as the correlation function calculated from the generated noise. The two lines overlap in all three cases. Panels (g) - (i) show (solid line) the population and (dashed line) the coherence in a homodimer. Parameters are $\sigma = 3.0 J$ and $\tau = 1.0 / J$. Time is in units of $1/J$.}
\end{figure*}

We now turn to the colored noise describing the effect of underdamped vibrations on the system, as defined in Eq.~(\ref{cf}). The effective damping rate in this case is given by
\begin{equation} \label{gammaeff}
  \Gamma_\mathrm{eff} = \frac{\Gamma_0}{1+\omega^2 \tau^2},
\end{equation}
which decreases with increasing $\omega$ for fixed $\Gamma_0$ and $\tau$. Thus, we expect that an overdamped dynamics for low vibrational frequency will change to an underdamped behavior for large $\omega$. This effect is shown in Fig.~\ref{fig:dimerseqcfpop}, which summarizes the simulation results in a homodimer ($\Delta=0$). Figs.~\ref{fig:dimerseqcfpop}(a) - (c) show a single realization of the noise for different values of the vibrational frequency $\omega$. The noise leads to fluctuations in the transition energy $\epsilon$, which is plotted relative to the amplitude of the noise $\sigma^2 = \Gamma_0/\tau$. In all panels the correlation time of the noise is $\tau = 1/J$.
Evidently, the colored noise caused by an underdamped vibration leads to oscillations in the transition energy, with a characteristic frequency given by $\omega$. The noise sequences are generated from the correlation functions shown in Figs.~\ref{fig:dimerseqcfpop}(d) - (f), which are typical for an overdamped vibration (Fig.~\ref{fig:dimerseqcfpop}(d)) and underdamped vibrations (Figs.~\ref{fig:dimerseqcfpop}(e) - (f)). The bottom three panels (Figs.~\ref{fig:dimerseqcfpop}(g)-(i)) show the population $P=\rho_{11}$ and the real part of the coherence $C=\mathrm{Re} [\rho_{12}]$ as a function of time, starting from an initial state where only molecule $1$ is populated.
Although the amplitude of the fluctuations $\sigma^2=\Gamma_0/\tau$ is the same in all panels, the effective damping strength $\Gamma_{\mathrm{eff}}$ in Eq.~(\ref{gammaeff}) is reduced with increasing $\omega$, leading to coherent oscillations in the population.
The proper damping strength by tuning vibration frequency may optimize the transfer efficiency.
Hence, coherent behaviors of the system emerges even in the presence of strong noise, and optimal transfer will occur for a critical
vibrational frequency $\omega_\mathrm{critical}$.

\emph{Resonant transfer}--The crossover from overdamped to underdamped system dynamics is the first effect that optimizes transfer in the dimer system. Now we consider the second mechanism. We expect the transfer efficiency to increase when the environmental vibration is resonant with the energy gap in the system.
Thus, for our dimer, resonant transfer will occur near the vibrational frequency $\omega_\mathrm{res} = \sqrt{\Delta^2 + 4 J^2}$, and can drive transfer between the two sites in the system. This effect can be understood as the driving of system dynamics by an external field, here provided by the vibration, in analogy to optical driving with a laser. This mechanism has been put forward as a way to regenerate coherence in the system\cite{Chin.2013.nphys.9.113}, and as quantum resonance\cite{Kolli.2012.jcp.137.174109}. Here, we show that it also enhances population transfer through classical resonance.
Thus, there are two mechanisms that control optimal transfer in the system: critical damping controlled by $\Gamma_\mathrm{eff}$, which can lead to a crossover from overdamped to underdamped system dynamics, and resonant transfer defined by $\omega_{\mathrm{res}}$ which matches the excitonic gap.

\begin{figure}[t]
\includegraphics[width=8.5cm]{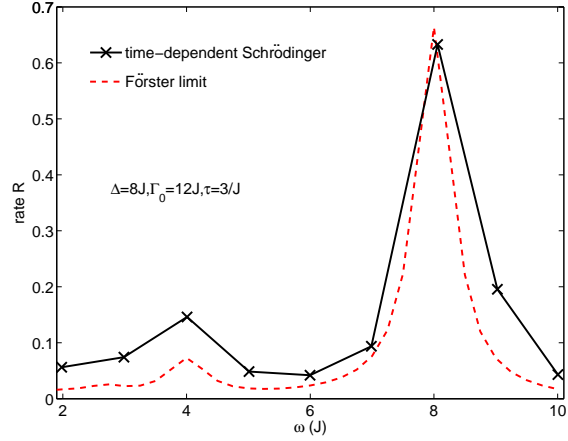}
\caption{\label{fig:twomaxima} Transfer rate from site 1 to site 2 as a function of vibrational frequency. Fixed parameters are $\Delta = 8 J$, $\Gamma_0 = 12 J$ and $\tau = 3 / J$. The two maxima in the transfer efficiency as a function of $\omega$, which correspond to the crossover from overdamped to underdamped dynamics ($\omega{\approx}4J$) and the exciton-vibrational resonance ($\omega{\approx}8J$), respectively.}
\end{figure}

The transfer rate between the two monomers can be understood from F\"{o}rster theory as follows. The corresponding F\"{o}rster rate in the presence of classical colored noise is given by\cite{Cao.2000.jcp.112.6719,Wu.2013.jcp.139.044102}
\begin{equation}
\kappa = 2 J^2 \mathrm{Re} [\int_0^\infty \mathrm{d}t e^{i\Delta t - g(t)}],
\end{equation}
where the line shape function is given by $g(t) = \frac{1}{2}\int_0^t \mathrm{d}t_1 \int_0^t \mathrm{d}t_2 L(t_1-t_2)$. For the correlation function given in Eq.~(\ref{cf}), the line-shape function is evaluated as
\begin{eqnarray} \label{Frate}
g(t)&=& \frac{\Gamma_0}{\tau(1/\tau^2+\omega^2)^2}\{\frac{t}{\tau}(1/\tau^2+\omega^2) \nonumber \\
    &+& (\omega^2-1/\tau^2)(1-e^{-t/\tau}\cos\omega{t}) \nonumber \\
    &-& \frac{2\omega}{\tau}{e^{-t/\tau}}\sin\omega{t}\}.
\end{eqnarray}
Hence, the transfer rate $R$, of which the inverse is defined as
\begin{equation} \label{ratedef}
R^{-1} = \int_0^\infty \mathrm{d}t (\frac{1}{2} - P(t))=\frac{1}{4\kappa},
\end{equation}
is readily evaluated,
where $P(t)$ is the transferred population at time $t$.
The result for the model dimer is shown in Fig.~\ref{fig:twomaxima}, where we indeed observe two peaks in the transfer rate as a function of vibrational frequency.

We now show how these two effects appear in numerical simulations, which can be applied to treat noise strengths outside the F\"orster regime.
From the transfer rate at Eq.~(\ref{ratedef}), the overdamped behavior is observed in low frequency regime.
By increasing $\omega$, there exists the crossover from overdamped to underdamped behavior, leading to optimal transfer for the critical value of $\omega_{\mathrm{critical}}$.
Thus, the prediction of Eq.~(\ref{gammaeff}) of a decrease of the effective noise strength with increasing $\omega$ holds.
However, the numerical solution of the vibrational frequency that leads to critical damping is different from the simple estimate, based on white noise result $\Gamma_\mathrm{critical} = \sqrt{\Delta^2 + 4 J^2}$. We attribute the difference to non-Markovian effects.
Surprisingly, the peak in the population transfer is rather sharp, as opposed to the broad peak observed for white noise.
If we keep increasing $\omega$, a resonance of the exciton system with the vibration appears at $\omega {\approx}8 J$.
Thus, the two maxima in the transfer efficiency as a function of $\omega$ correspond to the crossover from overdamped to underdamped dynamics, and to the exciton-vibration resonance.
Moreover, the strong increase of the transfer rate in the presence of a resonant vibration is found when the system is underdamped. In the overdamped case, it is confirmed there is almost no effect.
If $\omega$ is increased even further, the system dynamics becomes more coherent and the efficiency decreases dramatically.
In addition to the results presented here, we have performed simulations with the hierarchical equations of motion method for a Brownian oscillator spectral density, which models a quantized vibration. As shown in the Supplemental Material, we again observe the two maxima in the transport as a function of vibrational frequency.

\begin{figure}[t]
 \includegraphics{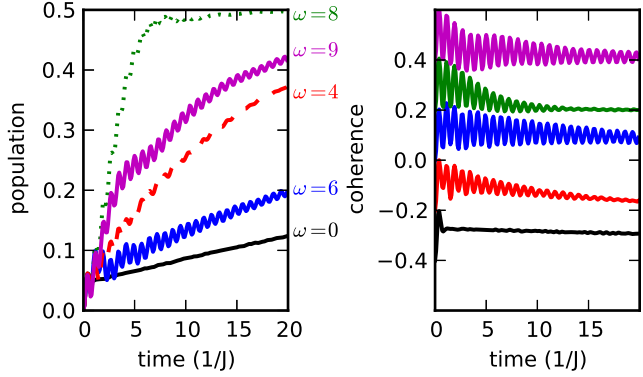}
\caption{\label{fig:coh} Population (left panel) and coherence (right panel) in the site basis as a function of time for different vibrational frequencies. Coherences were offset by -0.4 ($\omega=0$), -0.2 ($\omega=4 J$), 0 ($\omega=6 J$), +0.2 ($\omega=8 J$) and +0.4 ($\omega=9J$). The other parameters are identical to those in Fig.~\ref{fig:twomaxima}.}
\end{figure}

\emph{Suppressed coherence at resonance}--
Here, we also consider the quantum coherence between the two monomers in the site basis. We see long-lived coherent oscillations as a result of the coupling to the vibration.
This is a vibrational effect different from the absence of sustained oscillations at $\omega=0$. At resonance (i.e., $\omega_\mathrm{res} = 8J$), the coherence is damped faster than off-resonance. A similar faster decay is observed at $\omega_\mathrm{critical} = 4J$ and both effects can be
implied by more rapid population transfer. Thus, to observe long-lived coherence resulting from coupling to vibrations, one should probe away from the vibrational resonance.

The two mechanisms have different effects on the coherence of the dimer system.
The critical damping is the boundary between weak and strong damping and therefore cannot be described with either a weak or strong coupling master equation alone. For example, the standard Markovian version of the Redfield equation fails to predict the optimal energy transfer\cite{Plenio.2008.njp.10.113019}, whereas a non-Markovian version,
i.e., the generalized Bloch-Redfield equation, correctly predicts the transfer time and efficiency in the Fenna-Matthews-Olson complex (FMO) over the entire parameter space\cite{Wu.2010.njp.12.105012}.
In comparison, resonant transfer occurs at a weaker coupling strength and thus the standard Bloch equation is applicable. Then, in the resonant transfer regime,  the Bloch equation predicts $T_2=2T_1$, where $T_2$ and $T_1$ are the dephasing and population relaxation rate constants, respectively.  This simple relationship explains the rapid decay of excitonic coherence for resonant transfer.

\begin{figure}[t]
 \includegraphics{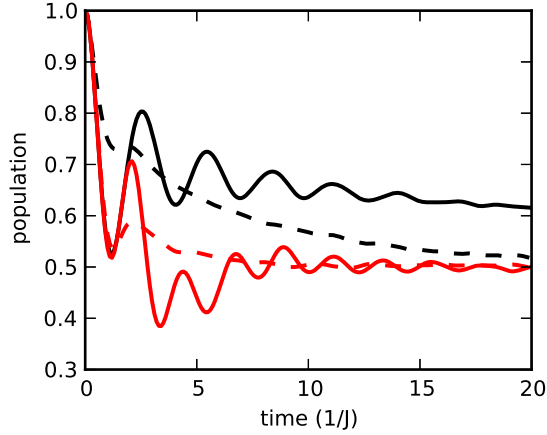}
\caption{\label{fig:optimal} Population on site 1 in a heterodimer as a function of time. Black lines are for exponentially correlated noise ($\omega = 0$), with $\Gamma_0 = 6 J$ for solid line and $\Gamma_0 = 96J$ for dashed line. Solid red line is for oscillating decay noise($\omega = 2.8J$), with $\Gamma_0 = 6J$. Dashed red line is for oscillating decay noise combined with exponentially correlated one shown at Eq.~(\ref{c2t}),
with $\Gamma_0=6J$, $\tau_2=2/J$ and $\Gamma_{0,2}=2J$.
The other parameters are $\Delta = 2J$ and $\tau=6/J$.}
\end{figure}

\emph{Combined effect and relevant photosynthetic systems}--We now show that the combination of the two optimization mechanisms leads to optimal transfer. Fig.~\ref{fig:optimal} shows the population in the heterodimer for exponentially correlated noise ($\omega=0$) in the underdamped regime (black solid line) and close to critical damping regime (black dashed line), respectively. In comparison, we also show the dynamics in the presence of a resonant vibration (red solid line). Finally, we plot the system population in the presence of both an underdamped and an overdamped vibration (red dashed line) by selecting the optimal parameters, with the combined correlation function
\begin{equation}~\label{c2t}
L_2(t) = \frac{\Gamma_0}{\tau} \cos (\omega t) e^{-|t|/\tau} + \frac{\Gamma_{0,2}}{\tau_2} e^{-|t|/\tau_2}.
\end{equation}
We observe that the latest case, in which a resonant vibration is present and the system is tuned close to critical damping, gives more efficient transfer, compared to other cases. Because the estimate of the effective damping in Eq.~(\ref{gammaeff}) does not quantitatively predict the crossover from overdamped to underdamped dynamics in the system, only numerical simulation can predict the parameters that may lead to maximum transfer efficiency. The combination of the two mechanisms identified in this paper, resulting in a resonant frequency $\omega_\mathrm{res}$ and a critical frequency $\omega_\mathrm{critical}$, leads to optimal transfer.

Recent experiments have detected oscillations of coherences in the heterodimer, 
which were attributed to both electronic coherence and vibrational states.
Many vibrational modes with frequencies close to the electronic energy gap were identified. Therefore, vibrational resonance effects are expected to play a central role in the population dynamics in these systems. A recent theoretical study considers an effective dimer model for the FMO complex\cite{Chin.2013.nphys.9.113}. The heterodimer has an energy offset of $\Delta = 2.43 J$. The spectral density considered consists of a broad-band with a reorganization energy of $\lambda_2 = 0.65 J$ and a vibration with a frequency of $\omega=3.36 J$ and reorganization energy of $\lambda=0.74 J$. Because in the high temperature limit $\Gamma_0/\tau = 2 \lambda/\beta$, we find that $\Gamma_0/\tau = 1.5 J^2$ and $\Gamma_{0,2}/\tau_2=1.3 J^2$ at a temperature of 77 K. At this temperature, the parameters are close to the optimal values used in our Fig.~\ref{fig:optimal}, suggesting that our simulations are directly relevant to real systems.
Further experiments that detect site populations are needed to show the
optimal transfer efficiency explained in the present paper.

\emph{Conclusion}--We have studied energy transfer in an exciton system coupled to an underdamped vibration.
The essential feature of the underdamped vibration can be obtained by considering intramolecular interaction.
Two mechanisms determine optimal transfer parameters.
First, there is a crossover from overdamped to underdamped system dynamics, governed by the effective strength of the noise. Because the strength of the noise depends on both the vibrational frequency as well as its damping constant, the behavior of the system changes with the frequency. Furthermore, a crossover from underdamped to overdamped dynamics can be tuned by changing the damping constant. This shows that experimentally controllable parameters, such as the choice of solvent, can modify the transport efficency, and even change the character of the population transport qualitatively.
Second, a vibration resonant with the excitonic gap can drive population transfer.
We found that resonance can strongly increase transport if the system dynamics is underdamped,
even in the presence of strong noise.
Optimal transfer is achieved when these two mechanisms coincide. Coupling to a vibration also leads to long-lived oscillations in the coherence,
which have the longest life time for a vibration that is slightly off-resonance.
Our analysis can trivially be applied to larger systems and to arbitrary correlation functions, including correlated fluctuations\cite{Wu.2010.njp.12.105012,Dijkstra.2010.njp.12.055005}.
Our results contribute to the heated debate on the effect of a vibrational resonance on energy transport.

\section*{Supplemental Material}
In the supplemental material, we present the derivation of the F\"orster rate shown in Eq.~(\ref{ratedef}) in the main text. Second, we present a plot of the transferred population as a function of time, in which the crossover from underdamped to overdamped behavior is visible. Finally, we present hierarchical equations of motion calculations for a quantized vibration.

\subsection{Derivation of transfer rate}

It is known that the transfer rate in the F\"{o}rster limit is given by
\begin{eqnarray}
\kappa=2J^2\textrm{Re}[\int^{\infty}_0d{t}e^{i{\Delta}t-g(t)}],
\end{eqnarray}
where $J$ is the inter-site coupling strength, $\Delta$ is energy splitting, and
the correlation phase is
\begin{eqnarray}
g(t)=\frac{1}{2}\int^t_0d{t_1}\int^t_0dt_2C(t_1-t_2)=\int^t_0d{t_1}(t-t_1)L(t_1),
\end{eqnarray}
with $L(t_1)$ the correlation function of the phonon bath.
In the high-temperature limit, the correlation function $L(t)$ is described by the colored noise
\begin{eqnarray}
L(t)=L_0(t)e^{-|{t}|/\tau},
\end{eqnarray}
with $L_0(t)$ the time dependent strength, and $\tau$ the characteristic correlation time.
Hence, the corresponding accumulation function is
$I(\tau,t)=\int^t_0d{t_1}L(t_1)$.
Then, the correlation phase is given by
\begin{eqnarray}
g(t)=(t+\frac{\partial}{\partial{r}})\int^t_0L(t_1)dt_1=(t+\frac{\partial}{\partial{r}})I(\tau,t),
\end{eqnarray}
with $r=1/\tau$.
If we further specify the correlation function as $L(t)=\frac{\Gamma_0}{\tau}e^{-|t|/\tau}$, with $\Gamma_0/\tau$ the fluctuation strength.
the accumulation function becomes
$I(\tau,t)={\Gamma_0}(1-e^{-t/\tau})$.
And the correlation phase is
\begin{eqnarray}
g(t)={\Gamma_0}{\tau}(e^{-t/\tau}-1+t/\tau).
\end{eqnarray}

In the strong damping limit $\tau{\rightarrow}0 (r{\rightarrow}\infty)$, the correlation phase is simplified to $g(t)=\Gamma_0t$.
The corresponding transfer rate is given by
\begin{eqnarray}
\kappa=2J^2\textrm{Re}[\int^{\infty}_0d{t}e^{i\Delta{t}-\Gamma_0t}]=2\frac{J^2\Gamma_0}{\Delta^2+\Gamma_0^2}.
\end{eqnarray}

If we consider a damped oscillation in the correlation function corresponding to intramolecular vibrations, the correlation function is modified as
\begin{eqnarray}
L(t)=\frac{\Gamma_0}{\tau}\cos(\omega{t})e^{-|t|/\tau},
\end{eqnarray}
with $\omega$ the vibrational frequency of the local mode.
The accumulation function is given by
\begin{eqnarray}
I(\tau,t)&=&\frac{\Gamma_0}{\tau}\int^t_0\cos(\omega{t_1})e^{-t_1/\tau}dt_1=\frac{\Gamma_0}{\tau}\textrm{Re}[\int^t_0d{t_1}e^{-i\omega{t_1}-t_1/\tau}]
=\frac{\Gamma_0}{\tau}\textrm{Re}[\frac{1-e^{-t/\tau-i\omega{t}}}{1/\tau+i\omega}]\nonumber\\
&=&\frac{\Gamma}{1/\tau^2+\omega^2}[1/\tau-e^{-t/\tau}(\frac{\cos\omega{t}}{\tau}-\omega\sin\omega{t})]
\end{eqnarray}
and the correlation phase is given by
\begin{eqnarray}
g(t)&=&(t+\frac{\partial}{{\partial}r})I(\tau,t)\nonumber\\
&=&\frac{\Gamma_0}{\tau(1/\tau^2+\omega^2)^2}[\frac{t}{\tau}(1/\tau^2+\omega^2)+(\omega^2-1/\tau^2)(1-e^{-t/\tau}\cos\omega{t})-\frac{2\omega}{\tau}{e^{-t/\tau}}\sin\omega{t}].
\end{eqnarray}

In the F\"orster limit, the dynamics of the populations are decoupled from the coherence term, and the former is expressed as
\begin{eqnarray}
\frac{dP_1(t)}{dt}=-\kappa(P_1(t)-P_2(t))=\kappa-2\kappa{P_1(t)}
\end{eqnarray}
with $P_j(t)$ the population at site $j$. This results in
$P_1(t)=\frac{1}{2}(1-e^{-2{\kappa}t})+P_1(0)e^{-2{\kappa}t}$.
If the initial condition is given as $P_1(0)=1$, the population dynamics of site $1$ are
$P_1(t)=\frac{1}{2}(1+e^{-2{\kappa}t})$.
Hence, the transferred population after a time $t$ is given by
$P(t)=1-P_1(t)=\frac{1}{2}(1-e^{-2{\kappa}t})$.
The inverse of the corresponding transfer rate is given by
\begin{eqnarray}
R^{-1}=\int^{\infty}_0d{t}(1/2-P(t))=1/(4\kappa),
\end{eqnarray}
and $R=4\kappa$.

\subsection{Hierarchical equations of motion results}

In order to show explicitly that the two peak structure found in this paper is still valid in a quantum mechanical calculation, we performed numerically exact simulations with the hierarchical equations of motion for the Brownian oscillator spectral density~\cite{Tanaka.2009.jpsj.78.073802}.
In this model, the energy difference between the two molecules is coupled to a vibrational mode, which is in turn coupled to a continuous bath of harmonic oscillators. The coordinates of the bath $X$ are treated as operators instead of functions of time. The spectral density for this model is given by
\begin{equation}
  J(\omega') = 2 \lambda \frac{\gamma \omega^2 \omega'}{(\omega^2-\omega'^2)^2 + \gamma^2 \omega'^2}.
\end{equation}
The parameters are the reorganization energy $\lambda$, which quantifies the strength of the system vibration coupling, the vibrational frequency $\omega$ and the coupling strength of the underdamped vibration to the thermal bath $\gamma$. The thermal bath is furthermore characterized by an inverse temperature $\beta$. Simulations were performed with one Matsubara frequency, a hierarchy depth of 6 and time steps of $0.001/J$.

\begin{figure}[t]
 \includegraphics{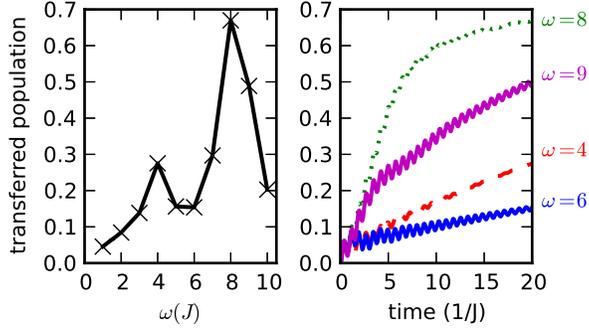}
\caption{\label{fig:bo} Transferred population from site 1 to site 2 (a) as a function of vibrational frequency after a time $20/J$ and (b) as a function of time for frequencies (from bottom to top) $\omega = 6J, 4J, 9J, 8J$. The dynamics were calculated with a quantized vibration. Fixed parameters are $\Delta = 8 J$, $\lambda = 0.2 J$, $\gamma = 2 J/3$ and $\beta=0.1/J$. The two maxima found in Fig.~\ref{fig:twomaxima} are reproduced in this calculation.}
\end{figure}

The results of these simulations are shown in Fig.~\ref{fig:bo}. We choose a temperature of $10 J$ and set $\lambda = 0.2 J$ to reproduce the same fluctuation amplitude as in the classical calculation shown in Fig.~\ref{fig:twomaxima} (In the high temperature limit $\lambda = \beta \Gamma_0/2 \tau$.). Although the amplitude depends on $\omega$, this choice of parameters reproduces a value of $\Gamma_0=12-13 J$ for all values of $\omega$ used. The Brownian oscillator spectral density gives rise to fluctuations damped with a characteristic time scale $\tau = 2/\gamma$, so we set $\gamma = 2 J/3$. Indeed, the two maxima in the transferred population are reproduced, showing the value of our simplified approach in predicting the behavior of the system. In the right panel of Fig.~\ref{fig:bo} the effect of the finite temperature is seen. At long times, the site populations reach thermal equilibrium instead of the value of $0.5$ found in the classical calculation reported in Fig.~\ref{fig:coh}. Essentially, the same effects are also found at much lower temperature, though not shown here.

\section*{Acknowledgements}
We thank T. Avila for carefully reading the manuscript. AGD thanks J. Moix and J. Cerrillo-Moreno for helpful discussions. The work at MIT was supported as part of the Center for Excitonics, an Energy Frontier Research Center funded by the U.S. Department of Energy, Office of Science, Basic Energy Sciences under Award \# DE-SC0001088. The work at LBNL and U.C. Berkeley was supported by the Director, Office of Science, Office of Basic Energy Sciences, of the USA Department of Energy under contract DE-AC02-05CH11231 and the Division of Chemical Sciences, Geosciences and Biosciences Division, Office of Basic Energy Sciences through grant DE-AC03-76SF000098.


\end{document}